# Differential Privacy in the 2020 Decennial Census and the Implications for Available Data Products

danah boyd / danah@datasociety.net
Principal Researcher, Microsoft Research and Founder/President, Data & Society
*Last updated July 8, 2019*

In early 2021, the US Census Bureau will begin releasing statistical tables based on the decennial census conducted in 2020. Because of significant changes in the data landscape, the Census Bureau is changing its approach to disclosure avoidance. The confidentiality of individuals represented "anonymously" in these statistical tables will be protected by a "formal privacy" technique that allows the Bureau to mathematically assess the risk of revealing information about individuals in the released statistical tables.[1] The Bureau's approach is an implementation of "differential privacy," and it gives a rigorously demonstrated guaranteed level of privacy protection that traditional methods of disclosure avoidance do not. Given the importance of the Census Bureau's statistical tables to democracy, resource allocation, justice, and research, confusion about what differential privacy is and how it might alter or eliminate data products has rippled through the community of its data users, namely: demographers, statisticians, and census advocates.

The purpose of this primer is to provide context to the Census Bureau's decision to use a technique based on differential privacy and to help data users and other census advocates who are struggling to understand what this mathematical tool is, why it matters, and how it will affect the Bureau's data products.

The Census Bureau's decision to implement differential privacy in the 2020 Decennial Census is borne out of its deep and statutorily mandated commitment to protecting the confidentiality of individual data records in order to maintain the trust of the public, whose cooperation and data the Census needs to do their work. The decision to proceed with implementing differential privacy now is driven by significant transformations in computational power and commercial data that make decennial data more vulnerable than in previous censuses. In short, research at the Census Bureau has shown that it is now possible to reconstruct information about and reidentify a sizeable number of people from publicly available statistical tables. As such, the leadership has accepted that they cannot continue with their current approach and wait until 2030 to make changes. In order to prevent reconstruction and reidentification from happening in the 2020 census, the Census Bureau is approaching disclosure avoidance in a fundamentally different way.

---

[1] Differential privacy is sometimes referred to as a type of "formal privacy" because it's mathematically provable (as in the proofs you did in high school geometry classes).

While differential privacy can and will protect the confidentiality of the data, it may also have significant costs. By taking this route, the Census Bureau may not be able to produce some of the data products that its data users have come to expect. Some data products will stay the same; others may not. As of this writing, the Census Bureau has not finalized which data products will and will not be available, prompting frustration among data users. To complicate matters more, the Census Bureau is effectively asking researchers, government agencies, and civil society to alter how they work with census data. For example, data users are being asked to articulate in advance what kinds of queries they intend to do in order to help the Census Bureau shape what products it should prioritize. This, too, has been met with confusion. This document will seek to clarify what is known – and what is not – as the 2020 decennial approaches.

**Background: The Commitment to Confidentiality**

Every ten years, the United States conducts a census of all people living in the country. The decennial census, which has been conducted every decade since 1790, is essential to representative apportionment and the fair distribution of shared financial resources. Participation in the census is required by law to uphold the Constitutional mandate of counting everyone. The decennial census was once conducted solely through door-to-door enumeration, but since 1970, there has been a self-response process in the form of a paper questionnaire returned by mail.[2] The 2020 Census will be the first decennial census conducted with four potential means of participation: internet self-response, telephone, paper, and door-to-door enumeration.[3] The latter, known at the Bureau as "non-response follow-up", is conducted after the self-response phase to make certain that every household ("living quarters address") is counted.

The innovation of paper-based self-response emerged for multiple reasons. The data were believed to be more accurate when people filled out the forms in the privacy of their own homes rather than sharing information with a stranger at the door. This "modern" way of collecting data also increased convenience (and, perhaps trust). It was also believed to lower the cost of conducting the census compared to universal field enumeration.

Trust has always been important to the Census Bureau. After all, the data collected must be as accurate as possible, but it also has to be trusted by the public and by Congress. When the census becomes politicized, as it was after the 1920 Census, trust in the process weakens. Throughout the 20th century,[4] there were a number of incidents that undermined trust in the Census Bureau and the data itself. One example concerned the 1940 Census, where data were provided to government officials to help locate people who were deemed to be a threat to national sovereignty during World War II, namely Japanese-Americans.[5] The human rights atrocities that ensued were not connected back to the census until decades later. Although today's laws prohibit a similar violation of confidentiality, those who have witnessed governmental violations of individual rights are often wary of the strength of legal protections.

---

[2] In 1960, the paper census was sent out by mail but enumerators went door-to-door to collect the paper forms.
[3] There are also additional operations for specific subpopulations, such as those living in group quarters and the unhoused.
[4] For a detailed historical account of the census, read Margo Anderson's *The American Census: A Social History*.

The Census Bureau has also taken numerous steps to ensure confidentiality of the data and the data collection process. For example, one critical step occurred in 1954, when Congress enacted Title 13 of the US Code.[6] Title 13 sets forth details about the administration of the census, confidentiality of census data, and penalties for violating confidentiality.[7] In short, the Census Bureau must produce statistics for the nation, but personal information cannot be given to government agencies, law enforcement personnel, immigration services, or courts of law. Of particular note, 13 U.S.C. §9 states that "Neither the Secretary, nor any other officer or employee of the Department of Commerce or bureau or agency [may]…. make any publication whereby the data furnished by any particular establishment or individual under this title can be identified." All who work for the US Census Bureau are sworn to uphold the confidentiality of the data (per Title 13) for life, an oath that is taken very seriously in the halls of the Bureau, and are subject to strict penalties, including fines and imprisonment.

While confidentiality is managed by both legal and procedural protections, as well as bureaucratic and technical ones, a public's trust in an institution like the decennial census is not wholly determined by the steps taken to ensure confidentiality. Furthermore, the political climate during any given census shapes public perception. Although the Census Bureau is comprised almost exclusively of non-partisan professionals from a range of scientific disciplines, the Bureau is a part of the Department of Commerce, which is an agency of the executive branch. Depending on who is in political office – and how the public feels about that administration – the public may perceive Census Bureau actions in a particular light.

All official communications that involve departments of the executive branch pass through the appointed secretary of that department, including routine, but mandatory, federal register notices and any reports to Congress mandated by law (e.g., content and questions on the decennial censuses). Inside the Bureau, both the Director and the Associate Director for Communication are also political appointees, with the candidate for Director requiring Senate confirmation. While the Decennial Directorate is explicitly non-political, the communication surrounding a given census can and often does have a political tinge to it. In a highly polarized political climate, the dynamics and perception of confidentiality are often shaped more by politics than practice.

---

[5] For more details on the relationship between the Census Bureau and Japanese internment, see Christa Jones' 2017 memo "Original Sources and Research Concerning Census Bureau Efforts to Support Japanese Internment": https://ecommons.cornell.edu/handle/1813/66186
[6] https://www.govinfo.gov/content/pkg/USCODE-2007-title13/pdf/USCODE-2007-title13.pdf
[7] In effect, Title 13 formalized the confidentiality norms that were already in place at the Census Bureau. For a more detailed history, see: Anderson, Margo and William Seltzer. 2007. "Challenges to the Confidentiality of US Federal Statistics, 1910-1965." *Journal of Official Statistics* 23(1), 1-34.

**Census Data Products and User Expectations**

The Census Bureau has always worked to balance the needs and interests of data users alongside the confidentiality of the information collected, recognizing the importance of public trust in the data collection process. Fundamentally, the loss of trust (which easily could be spurred by the loss of confidentiality) would also undermine robust data collection, rendering the data products less accurate and, in some cases, useless.

After the census is conducted, the Bureau produces different types of public data products. One class of data products consists of "statistical tables" (or "summary products").[8] These tabular summaries have geographic codes that go down to the census block.[9] The other major product is the Public Use Microdata Samples (PUMS) file that consists of a sample of individual responses. These "microdata" files have always been edited to remove personally-identifiable information and limit most geographic detail[10]; they also have larger margins of errors and are more complicated to work with. Still, some people find these microdata useful for specific research efforts.

The statistical tables and microdata produced from the decennial census are used for many purposes. For example, states use one type of statistical table – the redistricting files ("PL 94-171") – to draw electoral districts.[11] Federal laws mandate the collection and release of decennial census data necessary to allocate funding to states and localities. Other federal, state, and local laws, such as the Voting Rights Act, also require federal statistics. Furthermore, many local governments, non-profit and civil society organizations, federal agencies, and researchers have grown accustomed to using data from the Bureau even in situations where there is no statutory requirement for these stakeholders to rely on these data (or for the Bureau to produce it). As an agency collecting data about and for the public, there is a long history of making these data as broadly accessible as possible.

The data products that the Census Bureau publishes about the US population and households have changed over time. Perhaps the most significant change started when the Bureau began differentiating between enumerated data, which requires the participation of every household, and sampled data, which only includes statistically representative subset of households. Through the 2000 Census, the sampled "long-form" and complete enumeration "short-form" information were collected once a decade on the decennial census. Starting with the 2010 Census, the short-form data are still collected via the decennial operation and the long-form data are collected as part of the American Community Survey (ACS), a survey that samples a portion of the population and takes place on a rolling basis throughout the decade. While the apportionment of representatives requires decennial data only, most other federal laws and data uses can – and often do – use data from the ACS, as well as annually updated population estimates built off of the decennial census.

---

[8] For the non-technical, imagine a lot of different Excel files where each cell reflects the number of people that fit the characteristics on the X and Y axis. For example, the file could tell you how many "white Hispanic females ages 0-5" lived at a certain block.

[9] In 2010, there were ~11 million census blocks with ~6.2 million of them being inhabited.

[10] PUMS uses a geography code known as "public-use microdata area (PUMA)" which must have a population of at least 100,000 people.

[11] Legal fights during the 1960s resulted in the Supreme Court interpreting the Constitution as saying that districts had to be roughly equal in population count. These redistricting files help those constructing electoral districts and they also provide information for those challenging the maps as they are drawn.

The Census Bureau releases aggregated and anonymized data products, but does not make the underlying data available to the public or other governmental agencies. Individuals named on a record (or their heir) can examine original records upon request. Otherwise, personally identifiable census information remains confidential for seventy-two years, after which the National Archives and Records Administration is to release all individual records of a given decennial to the public. Currently, the most recent census available for inspection is the 1940 Census; this data is freely available to anyone who registers for access. In the meantime, no non-Census Bureau employee – including employees of other governmental agencies, academic researchers, or member of the public – can access this information.[12]

Data users can access a range of census data products through numerous means. Many statistical tables are made available to the public, particularly through the American Factfinder system. Anyone can download and use these tables at no cost and without permission from the Census Bureau. Researchers who need data that are not publicly accessible can access additional data (including a range of statistical tables and microdata) through the Federal Statistical Research Data Centers (RDCs). These centers are administered by the Census Bureau and require researchers to pass background investigations, swear for life to uphold the confidentiality of the data, and justify the benefits of doing the analysis. Notably, researchers working at the Federal Statistical RDCs must get approval before releasing data.

To make access to public datasets easier to use, researchers and scholars often package pre-processed statistical tables and microdata to help less technically savvy data users. For example, the Institute for Social and Data Innovation at the University of Minnesota hosts IPUMS, which helps many data users access such data. IPUMS also requires researchers to apply for access, but there is no federal background check or project approval required to work with their data.

---

[12] Census Bureau employees can only access this data if they have an approved research project that is deemed beneficial for the statistical work of the agency. All such employees are obliged to uphold confidentiality per Title 13.

There is also significant data use among Census Bureau employees, but only a designated subset of these employees can access the original data that contains personally identifiable information – and only if they have approval based on a justifiable need.

In choosing what statistical tables and microdata to release publicly, the Census Bureau has relied on detailed knowledge about the data and an in-depth understanding of contemporary statistical techniques. For decades, the Census Bureau has made numerous decisions not to release certain statistical tables or microdata in situations where it could be possible to learn something about individual respondents' records or undermine confidentiality. Advances in statistics and computer science, alongside the growing availability of commercial data, have consistently pushed the Bureau to innovate.

**The Production of Census Data Products**

The data products the Census Bureau releases publicly have never been perfect. There are a range of errors in the census data that arise from the collection and processing of the data. The Census Bureau has worked to remedy many of these errors over the years.

First, as much as the Census Bureau does everything possible to count all persons living in the US, some people are not counted – and some people are counted more than once. The net undercount has disproportionately affected certain subpopulations, including communities of color, immigrants, and children under 5. Second, people make mistakes when they self-respond. For example, a census record may come back indicating that a 3-year-old is the mother of a 27-year-old. Third, people may also fail to answer certain questions or provide answers that are unlikely (such as marking all options for all questions), or fail to include all residents of the household.[13]

Before the information collected during the census is turned into statistical tables, there are many steps that the Census Bureau takes to increase the quality of the data. The Bureau begins by creating a single file called the Decennial Response File (DRF) which contains data from self-responses and enumerator reports. In order to identify and remedy problems in this data, the team turns to administrative records collected and maintained by other governmental agencies (e.g., birth records or tax records). While the Census Bureau does not share its individual-level data with other agencies, other agencies share data with the Census Bureau in order to help with its mission. The Census Bureau uses administrative records during the data collection phase to better understand who they may not have counted, but after data collection is complete, they also use administrative records to improve the quality of the data. This data helps with correcting errors, filling out missing information ("imputing"), and identifying duplicates, among other things.[14] This clean-up process happens at multiple stages, with different issues addressed in order to produce the necessary statistical tables.

---

[13] In tabulating same-sex couples, census researchers recognize that some percentage of respondents appear to mark their sex inaccurately, thereby creating a situation in which some opposite-sex couples are marked as same-sex couples. See: http://socialcapitalreview.org/wp-content/uploads/2012/05/sshs2010c.pdf

The first post-census data product produced by the Census Bureau is the apportionment counts for the 50 states. These are delivered to the President by December 31st of the census year and are the basis for the reapportionment of the US House of Representatives. The apportionment counts are derived from the Census Unedited File (CUF) after the record count and state geography are locked. In producing the CUF, the Census Bureau takes numerous steps to identify potential flaws in data collection and processing, most notably to eliminate duplicate records. These counts are raw numbers of people per US state, with no additional detail.

In advance of producing more detailed statistical tables, the Census Bureau tries to correct for mistakes and omissions in individual attributes to produce the CEF ("Census Edited File"). As part of this, Bureau staff seek to identify and correct glaring data collection errors, such as the 3-year-old mother of a 27-year-old. They also *impute* missing data. In other words, if the Census Bureau does not know someone's age, it tries to predict it based on other characteristics of the household and/or administrative data. If the Census Bureau doesn't know someone's race and they don't appear in administrative data, employees might try to predict it based on information about the household or neighborhood. For example, if two members of the household marked themselves as Asian and two children were listed as Asian, the third child for whom race wasn't given might be marked as Asian. There are many different types of imputation that the Census Bureau uses to produce the CEF, all informed by the quality standards enforced by the Office of Management and Budget through the Chief Statistician of the United States.[15] The Census Bureau has conducted extensive research to inform its imputation process and make it as robust as possible. The CEF is used to produce a range of statistical tables, including population counts, redistricting files, and other materials that data users might want.

Internally, the Census Bureau seeks to maximize the accuracy of its data. Yet, when it makes statistical tables available, it must grapple with a second commitment: confidentiality. Since 1970, the Census Bureau has used a range of disclosure avoidance techniques to protect respondent confidentiality in the anonymous statistical tables it releases.[16] One technique – used since 2000 – is called "swapping." Through this technique, uniquely identifying data were changed so that individual people or families couldn't be identified. Say, for example, that a specific census block had one Black family. In that case, people in the community could easily identify that family and, thus, the other information they provided. Instead of publishing files with such uniquely identifying information, the Census Bureau would swap that family (and perhaps other families) with a family in another block.[17] The theory with swapping was that one could never be sure if a census record reflects real data for real people in the block, or if it a swapped record. Swapping is one of multiple disclosure avoidance techniques that have been historically used, and one of the techniques that will be entirely replaced by differential privacy.

---

[14] Urban Institute has produced a report on "Administrative Records in the 2020 US Census" (2017, authors: Dave McClure, Robert Santos, and Shiva Kooragayala) that discusses administrative records in greater detail: https://www.urban.org/sites/default/files/publication/90446/census_ar_report.pdf

[15] There have also been legal cases concerning different aspects of imputation. For example, in *Utah v. Evans*, the Supreme Court grappled with imputation of count in the creation of the CUF.

[16] These techniques are discussed in detail at McKenna, Lori. 2018. "Disclosure Avoidance Techniques Use for the 1970 through 2010 Decennial Censuses of Population and Housing," Census Research and Methodology Directorate, U.S. Census Bureau, Washington DC.: https://www2.census.gov/ces/wp/2018/CES-WP-18-47.pdf

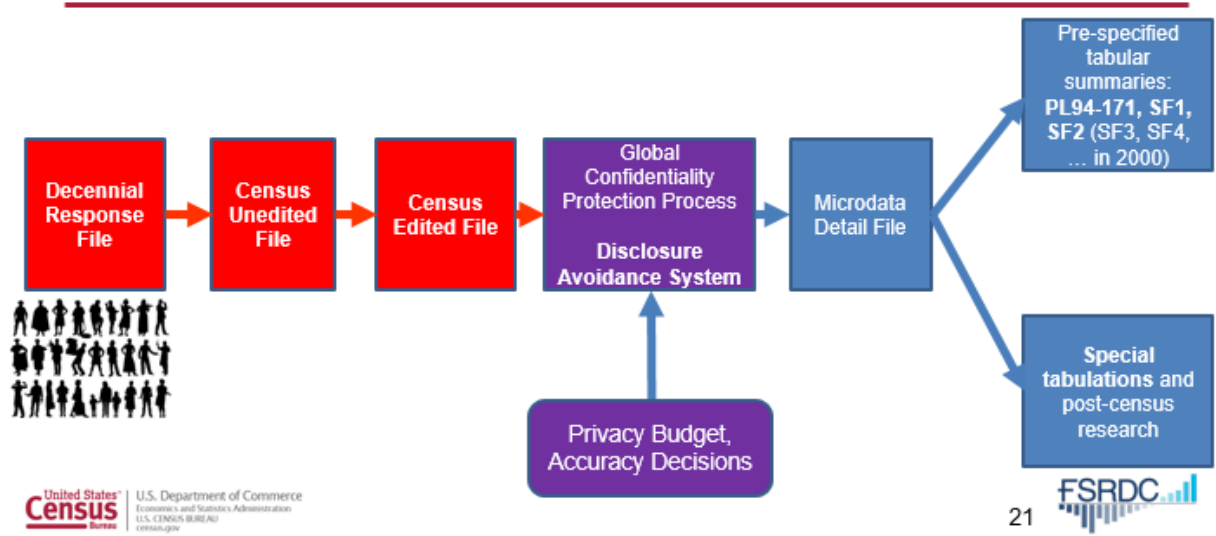
[18]

Disclosure avoidance techniques are designed to prevent breaches of confidentiality, but the level of "noise" to the system is not perceived as problematic, at least as far as statistical analyses go. Internal Census Bureau analyses suggested that the statistical error introduced through swapping was smaller than from other known sources of noise and error in the data

---

[17] This dynamic is documented in more detail in Mark Hansen's "To Reduce Privacy Risks, the Census Plans to Report Less Accurate Data," *New York Times*, December 5, 2018. In this story, he describes how the sole residents of Liberty Island (caretakers of the Statue of Liberty) were swapped before describing other aspects of differential privacy: https://www.nytimes.com/2018/12/05/upshot/to-reduce-privacy-risks-the-census-plans-to-report-less-accurate-data.html

[18] http://simson.net/ref/2018/2018-03-08%20Challenges%20and%20Experiences%20Adapting%20Differentially%20Private%20Mechanisms%20to%20the%202020%20Census%20(APPROVED).pptx

(e.g., respondent mistakes, non-response, editing, and imputation). Still, the Bureau has never released many details about how it conducted the swapping or what level of noise this technique produced as this information would have permitted external actors to reconstruct confidential information.[19] Moreover, statisticians and demographers have historically tolerated noise in the data, either because they understood how to manage accordingly or because they were unaware. There are also certain features that the Bureau has publicly stated are never touched; these are called "invariants." There are no statutorily required invariants, although the Bureau has long maintained certain categories as invariants (e.g., voting age population at all geographic levels, number of housing units at all geographic levels, etc.).[20]

In producing data products, there has always been a trade-off between accuracy and data availability. The introduction of swapping – and the presence of noise produced as a byproduct of data collection processes – has allowed the Bureau to maintain its promise of confidentiality while making relatively detailed data products widely available. Unfortunately, many people are unaware that noise exists, let alone how noisy some data are. For example, although block-level census data is highly prized, it is almost always so exceptionally noisy as to be meaningless for many types of analysis people use it for.

**Growing Challenges to Protecting Confidentiality**

Some data users are aware of the noise introduced by swapping, but they have never known the level of intentional blurring of the data in order to preserve confidentiality. However, in order to understand whether or not this noise was actually protecting the confidentiality of individual records, researchers began exploring whether or not they could reconstruct individual records using only the publicly available statistical tables. Before the 2010 census, academic researchers began using computational techniques to reconstruct individual records. Importantly, they developed techniques that could identify which records had been swapped, undermining the effectiveness of that disclosure avoidance strategy.

This research was part of a broader set of work on "reconstruction." In the late 1990s and early 2000s, privacy researchers, cryptographers, and security-minded computer scientists started raising concerns about the limitations of so-called "anonymous" open data. Recognizing that statistical tables "leaked" information about individuals, they began showing how it was possible to reconstruct individual records from statistical tables. Reconstruction allows an "attacker" to rebuild individual records from a set of statistical tables that are purportedly aggregates only.

---

[19] Even the Director of the Census Bureau is not provided with details on the swapping procedures. That said, the Census Bureau has confirmed that swaps never cross state borders.
[20] See: https://www2.census.gov/cac/sac/meetings/2019-03/managing-privacy-loss-budget-2020-census.pdf

In 2003, two researchers working in this space – Irit Dinur and Kobbi Nissim – devised a computationally efficient method for reconstructing individual information from statistical tables. In other words, they showed how a computer algorithm could easily do what was nearly impossible to do by hand. With answers to a large number of statistical questions, it was possible to accurately describe the responses in the confidential microdata from which a table was constructed. In other words, from statistical tables, individual responses could be discerned. Their work is known as the "database reconstruction" attack.

Reconstruction is a fancy way of describing a technique that allows computers to triangulate across different statistical tables (or data sources) to determine which individuals within the data are likely to contain which attributes. The greater the amount and detail of available data, the easier it is to transform a set of statistical tables into a list of individuals with known attributes, such as a person's age, sex, race, ethnicity, and the block on which they live.

Reconstruction does not allow the "attacker" to gain more information than was previously available in the aggregate data. If the statistical tables do not include a person's name, this method cannot provide that information. This is where a second attack comes into play. If an attacker has access to additional data that includes variables not present in the statistical tables, they can perform an "linkage attack" on reconstructed data. In other words, they can match individual records to external data to glean additional information. This is known as "reidentification."

For some people, reconstruction doesn't feel particularly invasive. After all, knowing that there exists a person who is White, Non-Hispanic, 17, and Female at a specific block may not mean much. But if this data can be matched with commercial data, it may be possible to know that this person is Jane Doe and learn her Facebook and Twitter accounts, along with countless additional information. Furthermore, if the census data contains attributes that people believe are sensitive (e.g., race, income, household makeup, citizenship), reidentifying the individual is seen as a breach of confidentiality.

Recognizing the potential risks of reconstruction, researchers at the Census Bureau attempted various reconstruction attacks in the run-up to the 2010 census. Based on what Census researchers knew from experts in the field and what they could assess themselves, they were confident that they could keep the 2010 data confidential. At the same time, they were concerned that this would not be true for decades to come. With this in mind, they began exploring different mechanisms to protect confidentiality and make various data products available.

**A Cryptographic Response**

In 2006, Cynthia Dwork, Frank McSherry, Kobbi Nissim, and Adam D. Smith proposed a new way to mathematically assess and guarantee the level of individual information that could be reconstructed from statistical tables. This method became known as "differential privacy" and was a mathematical definition for guaranteeing "privacy."[21] This concept opened up an entire field as researchers began looking for different ways to implement efficient systems for producing statistical tables that could be evaluated through the mathematical guarantees of differential privacy. In other words, because statistical tables "leak" information and because leakage is a vulnerability, these researchers developed a technique to evaluate how much leakage existed – and a process for evaluating how noise can be introduced to reduce the leakage.

Differential privacy works to prevent accurate reconstruction attacks while making certain that the data is still useful for statistical analyses. It does this by injecting uncertainty in the form of mathematical randomness, also called "noise", into the calculations that are used to produce the data products. With differential privacy it is still possible to reconstruct the database, but the database that is reconstructed will still include the privacy-ensuring noise. In other words, the individual records become synthetic byproducts of the statistical system. We will return to that later.

There are many different ways to build a system to protect the confidentiality of census data under differential privacy. In theory, a bottom-up way of applying differential privacy could be to alter the underlying data and then build the statistical tables from the altered data. In effect, this would be a form of swapping traits rather than swapping full individuals. Rather than having a 22-year-old, White, Non-Hispanic, Female and a 22-year-old, White, Hispanic, Male in the data, the individual records could be transformed to have a 21-year-old, White, Non-Hispanic, Male and a 23-year-old, White, Hispanic, Female. Certain statistical data would remain meaningful but individual entries would not, and the noise would be more evenly spread across the data, which would not be ideal for those seeking to answer questions about specific individuals.

Alternatively, a top-down approach might focus on the tables themselves. By understanding how data users might cross-tabulate variables or otherwise conduct statistical analyses, noise could be added to cells in the tables such that it would be difficult to discern individuals from statistical tables. To do this well requires understanding how different statistical tables interact with each other and would inevitably mean higher levels of noise in the tables with the most easily identifiable information (such as block-level data). But this would work because the attacker wouldn't know exactly how much noise was introduced into each cell. For example, a new table might say that there's two or even five 22-year-olds depending on how much noise is introduced into this part of the system.

---

[21] In discussions of differential privacy, computer scientists speak of "privacy" in a way that is similar to how census advocates speak of "confidentiality." What's at stake in terms of privacy in this context is whether or not you can discern features about individuals from aggregated data.

While both approaches are viable, the top-down approach can produce more tailored statistical tables because noise is injected in ways that preserve the most important statistical efforts. If race is the most important variable, more noise can be introduced to the age variable to preserve the accuracy of the race variable. If one table is more important than another table, more noise can be provided to the less-frequently-used table than the one that's more essential.

**Differential Privacy Controls**

The noise that is introduced in a differentially private system is not random in the colloquial sense. It is mathematically determined and relational, carefully chosen from a known distribution. As such, the noise is unpredictable for an attacker but constructed in a way that allows statistical analyses to work. The level of noise is defined by a variable known as *epsilon*. Epsilon is a mathematical variable that defines the privacy loss budget. Let's unpack this.

Every statistical table that is produced leaks some information. The more statistical tables the Census Bureau produces, the more likely that individual records can be accurately reconstructed. Each new table produced can be mathematically evaluated for how much new information it contributes to the total amount of information that an attacker could use. Noise can be introduced to the table to reduce the leaked information.

Within the construction of a differentially private system, it is necessary to make choices about how much noise is tolerable and how much risk (or "total privacy loss") is acceptable. The creators of a differentially private system set the maximum cumulative privacy loss budget. This is defined by the variable *epsilon*.[22] Epsilon can be conceptually understood as a knob (or, actually a set of knobs). Dial it one way and there are higher levels of noise but lower levels of confidentiality risk. Dial it the other way and the data are more accurate, but the risk to confidentiality (and, therefore, "privacy loss") is higher. The system has a global privacy loss budget, but each table also has a local privacy loss budget that must be managed such that the interaction of all tables does not result in leakage that exceeds the global amount.

Differential privacy requires a set of interconnected trade-offs. The lower the amount of noise injected into one particular table, the greater the accuracy of that table. Greater accuracy of one table means less accuracy is available for other tables because the total epsilon, or privacy-loss budget, is fixed. Once the privacy-loss budget is determined, the available accuracy must be shared among all the published tables. Increasing the accuracy of some tables without reducing the accuracy of others can only be accomplished by increasing the total privacy-loss budget, and therefore increasing the risk of confidentiality violations.

---

[22] Epsilon sounds fancy, but it's just a Greek letter. Mathematicians and computer scientists regularly use Greek letters as variables in their notation.

As a mathematical variable, setting epsilon requires making a choice between 0 and infinity, a decision that can be revised over time to reduce the noise if that becomes important.[23] De facto, in the case of the census, epsilon will be set to infinity at 72 years when the data are all made public. A more confidentiality-protecting choice in the global epsilon will be below 1, most likely around 0.25. Once this number is set, decisions need to be made about how to "spend" that budget based on the privacy loss of every statistical table that will be produced.

In effect, an implementation of differential privacy engenders two questions:
1) How should epsilon be set?
2) How should that privacy loss budget be allocated?

**From Exploration to Implementation**

The Census Bureau began exploring differential privacy as a potential avenue for addressing reconstruction attacks as early as 2006 and have implemented differentially private tables for select statistical products since 2008.[24] As a statistical agency filled with researchers who are advancing knowledge on many fronts, the Census Bureau consistently experiments with new statistical and computational techniques to advance their internal knowledge in the hopes that their findings can be helpful for future systems. Yet, in the case of differential privacy, Census researchers knew that they would need to make radical changes to the disclosure avoidance processes and this approach offered the most promising path forward.

In the run-up to the 2020 census, researchers at the Census Bureau decided to reassess the likelihood of a reconstruction attack being effective given technical advances. They also wanted to understand if commercial data might make reidentification more feasible. Using the publicly available statistical tables from only the 2010 decennial census, they could confidently reconstruct ~46% of the individual records of households in the 2010 decennial census. (If they relaxed the age to +/- 1 year, they could reconstruct ~71% of the individual records.) The accuracy of those reconstructions varied: some were almost certainly accurate, while others were much less likely to be accurate. Because Census Bureau researchers are in the unique position of being able to verify its reconstructions with original data, they were able to confirm that ~45% of the reconstructions were exact matches. With those reconstructions, the Census Bureau then attempted to match this data to commercial data. They were effectively able to match ~38% of the reconstructed decennial data to commercial data sources.[25] In other words, starting with the publicly available 2010 decennial statistical tables, Census Bureau researchers confirmed that they could reidentify ~17% of the public by name, a significant loss of confidentiality.

---

[23] While epsilon can be increased over time, it's not possible to decrease epsilon and increase privacy because the information has already been leaked.

[24] The Census Bureau was the first organization to release data protected by differential privacy. In 2008, they released the *On The Map* tool. See: https://www.census.gov/newsroom/blogs/research-matters/2018/08/protecting_the_confi.html

For the researchers inside the Census Bureau, their ability to perform this level of reconstruction and reidentification was a huge wake-up call. Recognizing how much commercial data had increased in the 2010s and, thus, how much more would be available by 2021, they had confidence that reidentification would only get easier. While external researchers might not be able to affirmatively verify the matches, Census Bureau researchers knew that many of the reconstruction and reidentification matches were overwhelmingly accurate and would only become more so.[26] In effect, it became clear to Census Bureau researchers that protecting the 2020 Census with the methods used in 2010 would yield a much higher reidentification rate than the 17% found with the external data available in 2010.

With the increased understanding that their 2010 approaches to protecting the confidentiality of census data would not be acceptable, the Census Bureau channeled significant energy into building a mechanism that could be guaranteed by differential privacy. By December 2018, the Census Bureau formally announced its intention to implement differential privacy. Unfortunately, this announcement was met with confusion on the part of many data users and census advocates. At the root of the problem was a disconnect in understanding about how statistical tables needed to be produced.

As part of their implementation of differential privacy, the Census Bureau must identify all of the statistical tables that it needs to publish and assess the total accuracy requirements of each one before production. Unlike the other techniques that the Census Bureau implemented in the past, critical decisions about the cumulative privacy loss budget must be decided in advance of releasing statistical tables to the public. Decisions concerning one table affect the production of future tables because the data are inter-dependent and the combination of tables is what leads to the loss of confidentiality.[27]

Once they know what tables are needed, the Census Bureau can then determine the maximum amount of noise that it can add while maintaining these requirements. If the Census Bureau was only concerned with producing the one constitutionally-required statistical table – the resident population of the 50 states – none of this would be necessary. However, additional statutes and regulations place demands on the privacy budget for the 2020 Census. The data used by the states for redistricting purposes are especially detailed with information reported at the block-level. In addition, there are a myriad of uses of the data by federal, state, and local agencies, as well as researchers and the business sector. In producing the statistical tables, the Census Bureau will try to address as many data users' needs as possible.

---

[25] https://www2.census.gov/programs-surveys/decennial/2020/resources/presentations-publications/2019-02-16-abowd-db-reconstruction.pdf
It is also important to note that commercial datasets have a lot more information about many people that the decennial census. Still, the ability to match decennial data to commercial data raises unique concerns, especially given Title 13.
[26] When a researcher knows that their match has a 99% likelihood, it is effectively "beyond a reasonable doubt" even if there is no formal confirmation of the match by the Census Bureau.
[27] Note to ACS data users: How the privacy budget is allocated on the decennial statistical tables has no bearing on the tables produced for the ACS. The only interdependence of these two surveys is at the sampling level, for which the CEF is used.

The Census Bureau has not yet solved the engineering challenges to produce differentially private versions of all the historically available public data products. As such, Census leadership is trying to navigate challenging trade-offs – while advancing the technical implementation – without disappointing too many data users. Census Bureau leadership must assess the privacy loss associated with the decennial census publications, and grapple with the trade-offs of determining which files are most important if the total cost would exceed the privacy loss budget. Only then can the Bureau produce the decennial census statistical tables.

When the Census Bureau announced their intention to implement differential privacy, they wanted the public to indicate what data products and fields were most important. Meanwhile, data users wanted to know which data products and fields would be available to assess whether or not they liked this change. These data users understood the Bureau as saying that fewer data products would be available, which was disconcerting to them; they did not understand that the Census Bureau's decisions would be dependent on their declarations of what specifically was needed.

The Census Bureau issued a public request for information in 2018 and was surprised that most people who submitted comments focused on data that is produced by the ACS, not the decennial census. Without knowing what tables people use and for what purposes, the Census Bureau has not been able to prioritize the availability of information or account for the level of noise that can be tolerated and how much of the privacy loss budget should be spent on a particular table. For example, if certain tables are not widely used, then there is really no reason to publish them or use up the budget on those datasets. But without adequate feedback from data users, the Census Bureau has had to guess.

Census data users are accustomed to unfettered access to tremendous amounts of public data. For this reason, people at the Census Bureau don't even know how different stakeholders use existing publicly available data, especially decennial data. They have never before needed to know who all of their users are or what they most need during the planning phases of census. Yet, in order to implement differential privacy and satisfy data users, they now need to know in advance what datasets are most important. This raises significant challenges about how to get this information and how to prioritize different needs, especially since some data users undoubtedly will be disappointed.

While all 2020 Census data products will be differentially private, the Census Bureau has not yet finalized which data products it will and will not be producing for the 2020 decennial.[28] This continues to frustrate and confuse many data users who are unfamiliar with how the Census Bureau must navigate the privacy loss budget trade-offs in making these decisions.

**Knowns and Unknowns Amidst Change**

As of this writing, we have yet to see exactly what the 2020 Census statistical files will look like, but we can make some strong assumptions based on the information that the Census Bureau has provided. For starters, the files that are released will have whole, positive numbers even though that attribute is not required for a differentially private dataset. Statistically, it's reasonable to say that 4.27 people live in a particular block. Yet, so as to not add even more confusion, the Bureau's technical team has chosen an approach that does not result in blocks with negative numbers of people or with partial people. This team has also focused on creating coherence among the files so that census blocks add up to tracts. This means that when noisy blocks are combined, they will produce less noisy tracts and so on up the geographic chain. In practice, this means that the redistricting file (also called the "P.L. 94-171 file") might say that there are 10 people living on one block, 20 people living on the next, and 5 people living on a third. In actuality, there might be 2 people on the first, 14 people on the second, and 18 on the third.[29] Although the individual blocks might have data that are quite inaccurate, when they are all added together, the larger geographies will be more accurate. In other words, because the Census Bureau's researchers have developed a "top-down" hierarchical algorithm, it is designed to produce statistics that are highly accurate at the county and state levels, but still protect confidentiality by concentrating the noise at the tract, block group, and block-level. They do this by defining the "workload" so that the most highly prioritized information (which has yet to be determined, but could be something such as state-level statistics or statistics representing a specific demographic) receives the least noise.

Second, the Census Bureau will continue to protect certain variables as "invariant." What this means is that there will be a small number of cells which represent exact enumerated counts, purposefully untouched by any noise. While there has been no official confirmation of all invariants, the Census Bureau has stated that the state-level population counts will be invariant. Although technically speaking, state-level counts could be differentially private and still be comparably accurate for apportionment, the political nature of these numbers is so great that protecting these cells as invariants is believed to be necessary. Invariants use up a greater proportion of the privacy loss budget so these decisions are extremely important as they significantly influence the amount of accuracy available for other parts of the system.

---

[28] At the 2019 Population Association of America Annual Conference, two representatives of the Population Division of the Census Bureau offered a proposal of what they intend to produce (including the apportionment product, the PL-94 redistricting file, a demographic profile, a demographic and housing characteristics file that would replace Summary File 1, and a congressional district demographic and housing characteristic file to replace the Census District Summary File). They also stated that, at this time, they are unable to produce detailed race and Hispanic origin tables and some household tables from Summary File 1, Summary File 2, American Indian and Alaska Native Summary File, and the Public Use Microdata Sample file.

[29] Note: In this example, one adds up to 35 and one adds up to 34. Depending on the implementation, it's possible that block-level deviation in numbers may not be permitted, but it will all depend on where it's acceptable to create noise in the data.

Unfortunately, differential privacy is likely to wreak the most havoc with data users who are not well-versed on the limitations of the data that they use. Many journalists, researchers, and policymakers have historically made claims using census data that were statistically inaccurate given the limitations of the data because they didn't understand (or perhaps even know about) those limitations. In providing detailed technical information, researchers at the Census Bureau are hoping to put pressure on data users to be more precise with how they construct their statistical analyses – or else their findings will be wildly off. In effect, one outcome of implementing differential privacy is that, by forcing questions to be articulated in advance, the Bureau is trying to eliminate one of the major practices of sloppy statistics.

Output produced using differential privacy can be used for statistical analyses. The statistics will hold for the calculations that were determined in advance, but it will not be possible to read specific subcategories as exact representations. In other words, if someone looks at a block-level file and sees that there are 6 Asian females over 65, they cannot – and should never have – assumed this to be accurate. It is important to note that this was never a valid way of reading the data, but most people were unaware of how much noise was introduced using previous methods; with differential privacy, the presence of noise will be much more explicit. Even if the exact amount of noise is not made available with every table, sophisticated data users will be able to calculate the margins of error from first principles. This is because the "random" noise will come technically from a known probability distribution.

Because systems that are designed to meet the specifications of differential privacy mathematically guarantee the upper bound of the "privacy loss," the details of these systems can be made available to the public. This is notable because, while the Census Bureau has long introduced noise into its statistical tables through swapping, it has never provided details about how much noise was injected. While the swapping process is easier to understand and describe, the implementation was never available to the public. For the 2020 census, the Census Bureau has already begun making its code and technical details publicly available and it will make decisions about the privacy loss budget, and its allocation to various tables, public too. It can do so because having this knowledge does not allow an attacker to undermine the system.

The Census Bureau is excited by the ability to be explicit and transparent about how the system is implemented, how much noise is introduced, and what can and cannot be meaningfully done with the data. The Census Bureau has already posted the source code that they are using. Even though differential privacy is more difficult to explain to the average person than swapping, technical data users will have greater ability to inspect the system. This creates a significant tension because the Bureau sees itself as increasing its transparency and providing *more* information, while many data users view what the Bureau is saying as being more confusing and less open because they do not yet know what tables will be made available and don't understand how differential privacy works. Moreover, non-technical data users and census advocates feel more confused than relieved by the availability of code.

**Anxieties and Uncertainties**

In trying to elicit information about how people use decennial data, Census Bureau researchers have discovered that many data users are not aware of what data they use and value. When Census Bureau employees explain how they are protecting decennial data, many data users and census advocates reasonably wonder what this effort will mean for the American Community Survey and other statistical products made available by the Census Bureau. The simple truth is: we don't know. But what we do know is that nothing will be changing anytime soon.

The team that is focused on building the differential privacy guarantees into their statistical table production process is currently focused exclusively on the statistical tables produced for the 2020 decennial census. Every aspect of how the ACS is conducted is different than the decennial. The ACS is a rolling statistical sample of the population, not an enumeration of the whole population's location on a single date. The statistical tables produced from the ACS are produced by weighting the sample data to reflect the whole population rather than using individual counts. Deliberate noise is introduced into the ACS through data editing in ways that makes reconstruction more difficult, although not impossible. ACS data and decennial data have never matched at the block-group level. Historically, because of swapping in the decennial and imputation in the ACS, the records that might be easiest to match between ACS and decennial due to their anomalous nature were likely to be protected by disclosure avoidance procedures. Even if an attacker developed techniques to match ACS against decennial data using previous methods, differentially private decennial data will make that impossible. Furthermore, the decisions to protect the confidentiality of decennial data have no direct impact on ACS data.

At present, Census Bureau researchers are still exploring how to reconstruct ACS data. They have not yet begun to evaluate whether or not – let alone how – to construct ACS tables using differential privacy. No matter what, the approach they take will need to be different than the approach they're taking for the decennial data. The sheer number of variables in the ACS, as compared to the decennial, fundamentally change the equation.

Another point of confusion concerns census microdata files (commonly known as PUMS, or Public Use Microdata Samples). PUMS data contains individual records for a subset of the population. PUMS have geographic area designations that must contain at least 100,000 people, unlike the block level geographic areas in the reconstructed microdata, which contain an average of 30 persons. PUMS files are also protected through disclosure avoidance techniques. Some data is simply not available, but that which is has historically been protected through the injection of noise. In other words, not all of the individuals in that data actually exist with those features.

In 2010, PUMS data was produced by taking a subset of individuals and then adding noise. Differential privacy requires a change to that approach. Rather than starting with accurate data and then injecting noise, the Census Bureau can start with the differentially private statistical tables and derive the individuals that would exist as a result of the noise.[30] In effect, both processes produce "synthetic" individuals who don't exist, but they are created in different ways. Through this process, synthetic records that represent no actual individual people are statistically equivalent to (if not better than) edited individual records with inconsistent noise, but this requires a more in-depth understanding of statistical methods than many data users actually have. By releasing PUMS, the Bureau runs the risk of increasing the misunderstanding of what microdata are able to say about individuals.

Finally, the Census Bureau also recognizes that some data users – namely, researchers – will need more access to better data than the Bureau can offer through the public data sets. Accordingly, Census leadership is seeking to expand the Federal Statistical Research Data Centers so that more researchers can receive Special Sworn Status and access additional data. The Bureau is also considering virtual programs that would eliminate the need to travel to an existing center, although none of these efforts will satisfy data users who cannot or do not wish to go through such programs. There is also notable overhead to participating in these programs – such data users must show that they have the statistical capabilities to work with these data, their research questions are evaluated, and they must get approval before publishing their results. Aside from the bureaucratic and organizational challenges this expansion demands, it also will create challenges vis-à-vis differential privacy. As researchers publish tables and statistical data from special access, the privacy loss budget must also account for the leakage of those data, and differential privacy will need to be applied to those tables, as well.

These changes and uncertainties have triggered anxiety in many people in the census world, from advocates to data users. They want clear information and are struggling to evaluate what differential privacy means for their work. This has been made more challenging by how few people understand differential privacy, and how those that do are often unable to communicate what they know in a way that resonates with non-technical audiences.

Very few people understand both how differential privacy transforms statistical tables and also how statisticians and demographers use census data. With the Census Bureau focused primarily on implementation and preparing example files, there is no one actively training data users about how to work with differentially private data. This has left a gaping hole in information that is being filled with anxiety. To complicate matters more, internal Census Bureau data users (e.g., demographers and statisticians) that rely on the decennial data are also trying to understand how the differentially private data might affect their work and the work of their external partners. Many of them do not understand the technical details of differential privacy, don't feel prepared to explain the technique to others in the field, and aren't in a position to assess the transformations like they have done in the past.

---

[30] While the Census Bureau officials have not made any final decisions about whether they will release differentially private microdata, they almost certainly will release the microdata file that undergirds the differentially private tables because they know it could be reconstructed.

**Taking Risk Amidst a Myriad of Challenges**

Differential privacy began as a theoretical computer science paper highlighting what was possible in the field of *privacy-preserving data analysis*. Slowly, companies started building systems that could provide differential privacy guarantees. Yet, even among the four largest implementors – Microsoft, Google, Apple, and Facebook – we have not seen an implementation that is as focused on making as much data available as possible. Fundamentally, the industry implementations are far more conservative because they have little incentive to make data accessible. By deciding to operationalize differential privacy in the 2020 decennial census, the Census Bureau has advanced scientific knowledge and exceeded the state of knowledge in industry in order to meet its public service mandate as a federal agency funded by taxpayer dollars.

By focusing on innovation and choosing to take the route of implementing differential privacy (rather than eliminating most data products), the Census Bureau is prioritizing data uses in ways that are poorly understood. Given the risks to confidentiality presented by the status quo, the choice is not between previous techniques and differential privacy. Rather, the choice is effectively differential privacy or bust. With no other technique available to guarantee confidentiality, the Bureau must either move forward with differential privacy or resist producing many constitutionally and statutorily required data products. As such, its focus is on how to spend the privacy budget optimally for the 2020 Census across a myriad uses where the only clear priorities are the constitutionally and statutory uses for reapportionment and redistricting.

The transformation of statistical tables through a differentially private system will have profound effects. First, it will reduce the accuracy of a reconstruction attack, which makes reidentification impossible. This allows the Census Bureau to assure that the reconstruction can't be used for purposes that would violate the protections of Title 13. But, in order to provide these protections, the Bureau will re-form the data products so that only certain analyses are meaningful. The data users must understand what can and cannot be done with the data if they want their analyses to be accurate, but we are not there yet.

As the Census Bureau team continues to plug away at the implementation for the Decennial, they are encountering numerous challenges. As mentioned earlier, they do not yet feel as though they have the information needed to determine which files and variables to prioritize when allocating noise. Furthermore, as the census draws nearer and more attention is paid to what the Census Bureau is doing, there's greater pressure to explain this technique to more people and train data users. Yet, with a stretched-thin team and a limited number of people outside of the Census Bureau who can train people, the Census Bureau is facing increased confusion among the growing number of census stakeholders.

More narrowly, the Census Bureau is implementing a system that opens up significant questions about governance and accountability. Who should be responsible for prioritizing what data is made publicly available? Who should be weighing the risks of confidentiality versus accuracy, setting epsilon, and allocating the privacy loss budget? And what oversight mechanisms should be in place to ensure appropriate decision-making in all of these areas? What processes should be put in place to make certain that the code is bug-free and that the calculations will hold at scale? Currently, there is an internal committee (Data Stewardship Executive Policy Committee) and an external committee (Census Scientific Advisory Committee) that are regularly briefed and consulted as part of the implementation process, but is this enough to account for all of the various data users? Now that data uses inform data production, how should the wide range of data users be involved in shaping the priorities of data production?

In conclusion, the Census Bureau is taking its pledge of confidentiality to the public — born of essential statutory requirements — with utmost seriousness. Census Bureau staff have taken an oath to protect the personal data of respondents. This commitment, in an era of "big data" where reconstruction and reidentification are increasingly possible, is challenging. Rather than compromising on confidentiality, the Bureau has taken the inspiring step of innovating. By integrating differential privacy into data production, the Census Bureau is pushing the boundaries of knowledge. While many details about the implementation of this technique are not yet known, one outcome is clear: the 2020 Decennial Census will maintain strict confidentiality, even if this limits the availability of public data.

**Acknowledgements**

This paper was made possible by spending countless hours with both differential privacy and census advocacy experts who have thoughtfully challenged me to better understand and communicate the Census Bureau's goals and plans. While not everyone appreciated public acknowledgement, I want to thank some of those who helped me improve this document along the way: John Abowd, Sareeta Amrute, Patrick Davison, Norin Dollard, Cynthia Dwork, Maria Filippelli, Abraham Flaxman, Kyla Fullenwider, Simson Garfinkel, Matt Goerzen, Daniel Goroff, Mark Hansen, Ron Jarmin, Charley Johnson, Jae June Lee, Terri Ann Lowenthal, William O'Hare, Ken Prewitt, Deborah Stein, Whitney Tucker, and Lars Vilhuber.